\documentclass[twocolumn,aps,prc,superscriptaddress,showpacs,nobibnotes]{revtex4}
\usepackage{epsfig,dcolumn,bm}

\newcommand{\dmu}{\partial_\mu}

\newcommand{\lc}{{\cal L}}

\newcommand{\sqd}{\sqrt{2}}

\newcommand{\be}{\begin{eqnarray}}
\newcommand{\ee}{\end{eqnarray}}
\newcommand{\nn}{\nonumber}

\begin{document}

\title{The radiative decay of $\psi(3770)$ into the predicted scalar state $X(3700)$}

\author{D. Gamermann}{\thanks{E-mail: daniel.gamermann@ific.uv.es}
\affiliation{Departamento de F\'isica Te\'orica and IFIC, Centro Mixto
Universidad de Valencia-CSIC,\\ Institutos de Investigaci\'on de
Paterna, Aptdo. 22085, 46071, Valencia, Spain} 
\author{E. Oset}{\thanks{E-mail: oset@ific.uv.es}
\affiliation{Departamento de F\'isica Te\'orica and IFIC, Centro Mixto
Universidad de Valencia-CSIC,\\ Institutos de Investigaci\'on de
Paterna, Aptdo. 22085, 46071, Valencia, Spain} 
\author{B.S. Zou}{\thanks{E-mail:zoubs@mail.ihep.ac.cn}
\affiliation{Institute of High Energy Physics and Theoretical Physics Center for Science Facilities, \\
CAS, Beijing 100049, China}
\affiliation{Departamento de F\'isica Te\'orica and IFIC, Centro Mixto
Universidad de Valencia-CSIC,\\ Institutos de Investigaci\'on de
Paterna, Aptdo. 22085, 46071, Valencia, Spain}

\begin{abstract}
We calculate the radiative decay width of the $\psi(3770)$ into the dynamically generated scalar resonance $X(3700)$ which is predicted in a previous paper. The results show that it is possible that the upgraded BEPC-II facility will generate enough statistics in order to observe this decay and thus confirm the existence of the $X(3700)$.
\end{abstract}



\maketitle

\section{Introduction}

The extension of $SU(3)$ Lagrangians to $SU(4)$, with the appropriate $SU(4)$ flavor symmetry breaking taken into account, has been proved useful in order to describe the structure of many charmed resonances as dynamically generated states \cite{meusca,meuax}. Although it presents a plausible explanation for many observed states, its spectrum shows many new states which have not been observed. The stability of some results, and the good fit achieved between the models and the observed experimental states, make us confident that the stable predicted states should exist. One should be careful with other states that seem to be more model dependent. For instance in \cite{lutz1,lutz2,guo1,guo2} two open charm narrow sextets, an axial and a scalar, are predicted while in our works \cite{meusca,meuax}, which take into account the differences between the meson decay constants of charmed and light mesons, these two sextets become very broad, the scalar one probably too broad to have any experimental relevance.

The model of \cite{meusca,meuax} also predicts a narrow hidden charm scalar resonance around 3.7 GeV and two axial vector resonances, nearly degenerated in mass, one with C-parity positive which corresponds to the $X(3872)$, and another one with negative C-parity. In order to stimulate the experimental search for these states, in the present work we present a calculation of the radiative decay width of the $\psi(3770)$ into the predicted scalar state around 3.7 GeV. We also discuss the possibility to observe this decay at the planned BEPC-II facility of Beijing based on the expected rate of $\psi(3770)$ production at this facility.

The work is organized as follows: In the next section we present the model for generating scalar resonances dynamically. We have done some improvements with respect to the work of \cite{meusca}, that we explain. Section III is dedicated to the framework to calculate the radiative decay width of vector mesons into scalar mesons and in section IV we point out our final remarks.

\section{Model}

First we build a field for the mesons belonging to the 15-plet of $SU(4)$, with help of the $SU(4)$ generators, $\lambda_i$:

\be
\Phi&=&\sum_{i=1}^{15}{\varphi_i \over \sqd}\lambda_i 
\ee
for this field a current is build:

\be
J_\mu&=&\dmu\Phi \Phi - \Phi\dmu\Phi \label{current}
\ee
note that adding a diagonal matrix (a $SU(4)$ singlet field) to $\Phi$ would give null contribution to $J_\mu$. In our previous work we have used only the mathematical states $\eta_8$ and $\eta_{15}$ that we have identified with the physical states $\eta$ and $\eta_c$ respectively. In the present work we are going to consider the mixing between these states and a singlet and use the matrix $\Phi$ in the physical basis:

\be
\Phi&=&\left(
\begin{array}{cccc}
 \frac{\eta }{\sqrt{3}}+\frac{\pi^0}{\sqrt{2}}+\frac{\eta'
   }{\sqrt{6}} & \pi ^+ & K^+ & \overline{D^0} \\
 \pi ^- & \frac{\eta }{\sqrt{3}}-\frac{\pi
   ^0}{\sqrt{2}}+\frac{\eta'}{\sqrt{6}} & K^0 & D^- \\
 K^- & \overline{K^0} & \sqrt{\frac{2}{3}} \eta'-\frac{\eta
   }{\sqrt{3}} &  {D_s}^- \\
 D^0 & D^+ &  {D_s}^+ & \eta _c
\end{array}
\right) \nn
\ee

The Lagrangian we use is the current in eq. (\ref{current}) coupled to itself plus a mass term:

\be
\lc={1\over12f^2}Tr({J}_\mu {J}^\mu+\Phi^4 M) \label{lag}.
\ee

For the mass term we take the matrix $M=diagonal(m_\pi^2,m_\pi^2,2m_K^2-m_\pi^2, 2m_D^2-m_\pi^2)$. This Lagrangian is a straightforward generalization to $SU(4)$ of the usual lowest order chiral Lagrangian used for $SU(3)$ in several works \cite{dipl1,sca1}. Since $SU(4)$ is not a good symmetry in nature we are going to explicitly break it in three ways. First, since the mass term is not proportional to the identity matrix, it already has some amount of $SU(4)$ symmetry breaking. The parameter $f$ in (\ref{lag}) is the meson decay constant, we are going to use $f=f_\pi=93$ MeV for light mesons, and $f=f_D=165$ MeV for heavy ones, and last we are going to suppress terms in the Lagrangian where the interaction is driven by the exchange of a heavy vector meson, in the vector meson dominance picture. For details in this suppression one should refer to \cite{meusca,meuax,hofmann,angels}.

From the resulting Lagrangian, using usual Feynman rules, the transition amplitudes between any possible two meson initial and final states that span a coupled channel space are calculated, projected in s-wave, and collected into a matrix, $V$. This matrix is then used as the kernel to solve the Bethe-Salpeter equation for the T-matrix, that in the on-shell formalism of \cite{noverd,oller}, assumes an algebraic form:

\be
T=V+VGT \label{bseq},
\ee
in this equation $G$ is a diagonal matrix with each one of its elements given by the loop function for each channel in the coupled channel space.

The imaginary part of the loop function ensures that the T-matrix is unitary, and since this imaginary part is known, it is possible to do an analytic continuation for going from the first Riemann sheet to the second one. Possible physical states (resonances) are identified as poles in the T-matrix calculated in the second Riemann sheet for the channels which have the threshold below the resonance mass.

Considering all the mesons in the 15-plet of $SU(4)$ plus the singlet field, it is possible to couple 15 channels to the quantum numbers C=0 and S=0: $\pi^+\pi^-$, $K^+K^-$, $\pi^0\eta$, $\pi^0\eta'$, $D^+D^-$, $\eta_c\pi^0$, $\pi^0\pi^0$, $K^0\overline{K^0}$, $\eta\eta$, $\eta\eta'$, $\eta'\eta'$, $D^0\overline{D^0}$, $D_s^+D_s^-$, $\eta_c\eta$ and $\eta_c\eta'$. Three of these channels are pure isospin 1 channels and, since as discussed in \cite{meusca} the hidden-charm scalar resonance is an isospin 0 resonance, it will not couple to those channels.

\begin{table}[h]
\begin{center}
\caption{Couplings of the pole at (3722-$i$18) MeV to the channels.} \label{tab1}
\begin{tabular}{c|c|c|c}
\hline
Channel&Re($g_X$) [MeV]&Im($g_X$) [MeV]&$|g_X|$ [MeV]\\
\hline
\hline
$\pi^+\pi^-$&9&83&84\\
\hline
$K^+K^-$&5&22&22\\
\hline
$D^+D^-$&5962&1695&6198\\
\hline
$\pi^0\pi^0$&6&83&84\\
\hline
$K^0\overline{K^0}$&5&22&22\\
\hline
$\eta\eta$&1023&242&1051\\
\hline
$\eta\eta'$&1680&368&1720\\
\hline
$\eta'\eta'$&922&-417&1012\\
\hline
$D^0\overline{D^0}$&5962&1695&6198\\
\hline
$D_s^+D_s^-$&5901&-869&5965\\
\hline
$\eta_c\eta$&518&659&838\\
\hline
$\eta_c\eta'$&405&9&405\\
\hline
\end{tabular}
\end{center}
\end{table}

Solving the coupled channels problem for calculating the T-matrix with the same parameters as in \cite{meusca}, but with the physical basis for the interacting fields (considering the $\eta$, $\eta'$ mixing and the $\eta_c$ as a pure $c\bar{c}$ state) we find now the $X(3700)$ pole at $\sqrt{s}=$(3722-$i$18) MeV. In Table \ref{tab1} we show the couplings of this resonance to all channels calculated via the residues of the pole to the channels.

Note that now the $X(3700)$ has a much bigger width than in \cite{meusca} where this width was smaller than 1 MeV. This is due to the fact that in \cite{meusca} the only decay channels which were not strongly suppressed by the dynamics of the interaction were channels involving only heavy mesons. The lighter channels were suppressed by the interaction and the channel $\eta_{15}\eta_{8}$, which is closer in mass to the resonance was suppressed because it is a pure $SU(3)$ octet while the $X$ is a $SU(3)$ singlet. Now, the $\eta$ and $\eta'$ have an important $SU(3)$ singlet component and so the couplings of the resonance to channels involving these mesons is bigger and therefore the resonance acquires a bigger width, although still small compared to other hadronic resonances.

In the next section we present the formalism to evaluate the radiative decay width of a vector into a scalar.

\section{Radiative Decay}

We are going to consider the following radiative decay: 

\be
\psi(P,\epsilon_\psi)\rightarrow X(Q)+\gamma(K,\epsilon_\gamma) \label{decay}
\ee
where $\psi$ is the second radial excitation of the charmonium spectrum, the $\psi(3770)$ and $X$ is the dynamically generated resonance $X(3700)$.

We are following here the same steps as in \cite{meurad} for the calculation of the radiative decay width, which is based in the formalism used in \cite{close,ollerrad}. In the reaction (\ref{decay}) one has two independent four momenta since $P=Q+K$. We chose to work with $P$ and $K$. Moreover both vector particles fulfill the Lorentz condition:

\be
P_\mu\epsilon_\psi^\mu&=&0 \label{lor1}\\
K_\mu\epsilon_\gamma^\mu&=&0 \label{lor2}
\ee

The amplitude for the decay in (\ref{decay}) will be given by

\be
-i{\cal M}&=&\epsilon_\psi^\mu\epsilon_\gamma^\nu{\cal T}_{\mu\nu},
\ee
and since the problem has two independent four momenta, by Lorentz invariance one may write

\be
{\cal T}_{\mu\nu}&=&ag_{\mu\nu}+bP_\mu P_\nu+cP_\mu K_\nu+dP_\nu K_\mu+eK_\mu K_\nu. \label{tmunu}
\ee
Now one can realize that by Lorentz condition (\ref{lor1}),(\ref{lor2}) the terms with coefficients $b$, $c$ and $e$ will not contribute to the radiative decay amplitude.

Applying the gauge condition ($K^\nu{\cal T_{\mu\nu}}=0$) to the expression in (\ref{tmunu}) one obtains that the two remaining coefficients are related: $a=-d K.P$, so one needs to calculate only one of them in order to obtain the full gauge invariant amplitude for the process. We are going to calculate the $d$ term, which comes from only one diagram, illustrated in figure \ref{fig1}. In the figure $m_1$ and $m_2$ are the masses of the charged mesons to which the $X(3700)$ can couple, we will consider only the $D^+D^-$ and $D_s^+D_s^-$ channels, since the other charged channels to which it couples have negligible couplings compared to these ones, see table \ref{tab1}.

\begin{figure}[t]
\begin{center}
\includegraphics[width=7.5cm,angle=-0]{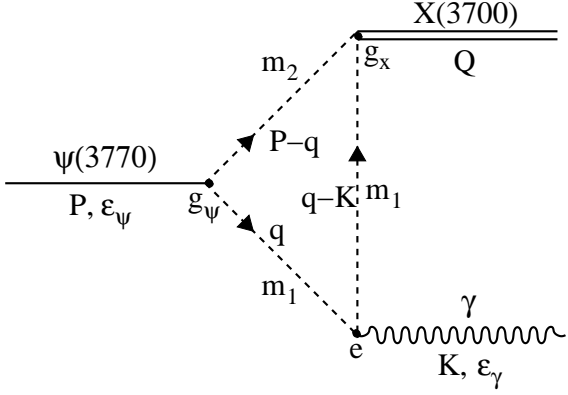} 
\caption{Diagram that contains the $d$ term.} \label{fig1}
\end{center}
\end{figure}

Given the basic couplings of $\psi(P)\rightarrow D^+(q)D^-(P-q)$:

\be
-i{\cal M}_{\psi\rightarrow D^+D^-}&=&-i g_\psi \epsilon_\psi^\mu (q-P+q)_\mu,
\ee
of $X(3700)$ to $PP$:

\be
-i{\cal M}_{X\rightarrow PP}&=&-ig_X,
\ee
and of the photon to $PP$:

\be
-i{\cal M}_{\gamma\rightarrow PP}&=&-ie \epsilon_\gamma^\nu(q+q-K)_\nu,
\ee
for a positive pseusdoscalar meson and $e$ being the modulus of the electric charge, the amplitude for this process is given by:

\be
-i{\cal M}&=&4g_\psi g_X e \epsilon_\psi^\mu \epsilon_\gamma^\nu \int {d^4q\over(2\pi)^4} q_\mu q_\nu  \nn \\
&\times& {1\over (P-q)^2-m_2^2} {1\over q^2-m_1^2 } {1\over(q-K)^2-m_1^2}
\ee

Using Feynman parameterization one gets:

\be
{1\over abc}&=&2\int_0^1dx\int_0^x dy {1\over \left(a+(b-a)x+(c-b)y\right)^3} \\
a&=&(P-q)^2-m_2^2 \\
b&=&q^2-m_1^2 \\
c&=&(q-K)^2-m_1^2 \\
{\cal T}_{\mu\nu}&=&8g_\psi g_X e \int_0^1dx\int_0^x dy \int {d^4q'\over (2\pi)^4} \nn \\
&\times&{(q'+yK)_\mu (q'+(1-x)P)_\nu 
 \over (q'^2+s+i\epsilon)^3 },\label{intintint}
\ee
where in the last equation we have made a change of variables $q=q'+(1-x)P+yK$ and we have defined $s$:

\be
s&=&(1-x)(x m_\psi^2-m_2^2-2yP.K)-x m_1^2
\ee

Now we can identify the $d$ term in the expression of equation (\ref{intintint}) and perform the $q'$ integral.

\be
\int {d^4q'\over (q'^2+s+i\epsilon)^3}&=&{i\pi^2\over 2}{1\over s+i\epsilon}
\ee

The coupling of the photon to the other particle ($m_2$ in figure \ref{fig1}) gives, in the present case, an identical contribution since $D^+D^-$ have opposite signs and equal mass and the same occurs for $D_s^+D_s^-$. Hence for each of these two channels we obtain:

\be
d=i{g_\psi g_X e\over 2\pi^2}\int_0^1dx\int_0^x dy {y(1-x)\over s+i\epsilon} \label{dterm}
\ee
and we must sum this expression for the two channels $D^+D^-$ and $D_s^+D_s^-$ using the coupling $g_\psi$ to each channel that we evaluate below, and $g_X$ which appears in Table \ref{tab1}.

With the $d$ term one can calculate the $a$ term and with these two terms the full gauge invariant amplitude $\cal M$ for the radiative decay. Making an average over the possible polarizations of the $\psi(3770)$ state of the absolute value square of this amplitude, summing over the possible photon polarizations and integrating over phase-space, one obtains the radiative decay width:

\be
\Gamma_{\psi\rightarrow X\gamma}&=&{p\over 12\pi m_\psi^2} (P.K)^2 |d|^2, \label{radwidth}
\ee
where $p$ is the relative center of mass three momentum of the two particles in the final state.

There is still one parameter to be calculated which is the coupling of the vector $\psi(3770)$ to the two possible channels.

At tree level the decay width of a vector meson into two pseudoscalars is given by:

\be
\Gamma_{V\rightarrow PP}&=&{1\over 8\pi M_V^2}{4 \over 3} g^2 p^3, \label{vdecay}
\ee
where $g$ is the $VPP$ coupling and $p$ is the center of mass relative momentum of the two pseudoscalars in the final state.

The total decay width of the $\psi(3770)$ is 25 MeV where 36\% of this width is coming from $D^+D^-$ final state, according to \cite{pdg}. Using 9 MeV for $\Gamma$ in eq. (\ref{vdecay}) one obtains 11.7 for the value of $g$. The $\psi(3770)$ is the $3S$ radial excitation of the $J/\psi$ vector meson. Hence we expect the flavor dynamics of this state to be identical to that of the $J/\psi$. This means that the couplings of $\psi(3770)$ to the different $PP$ states will be the same as those of the $J/\psi$ up to a global proportionality constant related to the overlap of the radial wave functions.

We follow the work of \cite{meusca} and write the $VPP$ Lagrangian in $SU(4)$ formalism:

\be
{\cal{L}}_{PPV}&=&\frac{-i g}{\sqrt{2}}Tr([\partial_\mu \Phi,\Phi]V^\mu)
\ee
with

\be
V_\mu&=& \left(
\begin{array}{cccc}
\frac{\omega }{\sqrt{2}}+\frac{\rho ^0}{\sqrt{2}} & \rho ^+ & K^{*+} & \overline{D^{*0}} \\
 \rho ^- & \frac{\omega }{\sqrt{2}}-\frac{\rho ^{*0}}{\sqrt{2}} & K^{*0} & D^{*-} \\
 K^{*-} & \overline{K^{*0}} & \phi  & D_s^{*-} \\
 D^{*0} & D^{*+} & D_s^{*+} & \text{J/$\psi $}
\end{array}
\right)_\mu \nn
\ee

This guarantees that the coupling of $\psi(3770)$ to $D^+D^-$ and to $D_s^+D_s^-$, which are needed in the loop of figure \ref{fig1}, are equal, like in the case of $J/\psi$.

With the expressions in eqs. (\ref{dterm}) and (\ref{radwidth}) it is straightforward to evaluate the radiative decay width of the $\psi(3770)$ into the $X(3700)$ resonance.

For the masses of the mesons we have used: $m_D$=1868 MeV, $m_{D_s}$=1965 MeV, $m_X$=3722 MeV and $m_\psi$=3772 MeV.

Applying the expressions in (\ref{dterm}) and (\ref{radwidth}) one obtains for the radiative decay width of the $\psi(3770)$ into the $X(3700)$:

\be
\Gamma_{\psi\rightarrow X\gamma}&=&0.65 {\rm \hspace{0.15cm} KeV}. \label{result}
\ee

Taking into account the finite width of the $\psi(3770)$ or of the $X(3700)$ in the evaluation of this radiative decay has less than 10\% effect in the result. 

The value in (\ref{result}) corresponds to a branching fraction of:

\be
{B(\psi(3770)\rightarrow X(3700)+\gamma)\over B(\psi(3770)\rightarrow {\rm anything})}&=&2.6\times 10^{-5}
\ee

The branching ratio is of the order of magnitude of the $\phi\rightarrow a_0(980)\gamma$ or $\phi\rightarrow f_0(980)\gamma$ decays \cite{phidec1,phidec2}, which proceed with similar loops but involving kaons instead of $D$ mesons \cite{phiteo1,phiteo2}.

It is interesting to make an estimative of the theoretical uncertainties. In \cite{ddbumps} it was suggested that the bump in the $D\bar D$ spectrum close to threshold observed at Belle in the $e^+e^-\rightarrow J/\psi D\bar D$ \cite{belle} could be due to the existence of the hidden charm state that we have considered. The alternative explanation as a new state above the $D\bar D$ threshold, mildly suggested in \cite{belle}, had a $\chi^2$ value larger than the one obtained assuming the peak as due to the hidden charm state below threshold. The best fit to the data was obtained with a mass of the $X(3700)$ around 3720 MeV. We have changed the subtraction constant of the loop function in order to generate the $X(3700)$ pole at different positions which still produced good $\chi^2$ values in the fit of the $e^+e^-\rightarrow J/\psi D\bar D$ data, and we get the results for the radiative decay width shown in Table \ref{tab2}.

\begin{table}[h]
\begin{center}
\caption{Results for different pole positions of the $X(3700)$ resonance. The value in the $E_\gamma$ column corresponds to the photon momentum.} \label{tab2}
\begin{tabular}{c|c|c|c|c|c}
\hline
$\alpha_H$&$\sqrt{s}_{pole}$ [MeV]&$g_D$ [GeV]&$g_{D_s}$ [GeV]&$E_\gamma$ [MeV]&$\Gamma$ [KeV]\\
\hline
\hline
-1.40&3702-i27&8.06+i1.29&7.83-i1.34&96.73&3.24 \\
\hline
-1.35&3713-i21&6.52+i1.54&6.39-i1.06&58.54&0.99 \\
\hline
-1.30&3722-i18&5.96+i1.69&5.90-i0.87&49.67&0.65 \\
\hline
-1.25&3730-i15&5.39+i1.91&5.42-i0.60&41.77&0.43 \\
\hline
\end{tabular}
\end{center}
\end{table}

The feasibility of the experiment at BES relies upon the statistics of $\psi(3770)$ production. According to \cite{ablikim,haibo} in BEPC-II it is expected to have $3.8\times 10^7$ $\psi(3770)$ events in one year of run, which would correspond to around 1000 events into $X(3700)\gamma$ for $\Gamma_{\psi\rightarrow X\gamma}$=0.65 KeV, with the photon energy peaking around 50 MeV. For other values of the $X(3700)$ mass the photon peak position changes and so does the rate, but in all cases disregarding technical problems that are beyond our reach, the statistics of one year is far more than sufficient to determine this peak with the required precision.

To further support our claim we would like to quote that CLEO \cite{cleo} has reported the observation of the $\psi(3770)$ decay into $\gamma \chi_{c0}(1P)$ which has a decay width of 172 KeV, about two orders of magnitude bigger than our results for $\gamma X(3700)$. Since BEPC-II will have two orders of magnitude more statistics than CLEO, the measurement that we suggest seams to be at reach.

\section{Overview}

The fact that the predicted hidden charm resonance $X(3700)$ lies below the $\psi(3770)$, together with the plans to produce this latter particle at BEPC-II, opens up the possibility to see the $X(3700)$ resonance as a narrow peak around 50 MeV in the photon spectrum from the radiative decay of $\psi(3770)$ into $X(3700)\gamma$.

The predicted radiative decay, obtained following an approach which has been successful in predictions of the radiative decay of $\phi$ into $a_0(980)$ and $f_0(980)$, should be rather reliable. Based on the calculated rate, the planned run of $\psi(3770)$ production at BEPC-II should produce enough events to gather sufficient statistics of decays into the $X(3700)\gamma$ channel. The study done here should stimulate experimental efforts in this direction.

\acknowledgments{One of us, D. Gamermann, wishes to acknowledge support from the Ministerio de Educacion y Ciencia in the FPI program. This work was supported in part by DGICYT contract number FIS2006-03438, the Generalitat Valenciana, the National Natural Science Foundation of China and the Chinese Academy of Sciences under project number KJCX3-SYW-N2. This research is part of the EU Integrated Infrastructure Initiative Hadron Physics Project under contract number RII3-CT-2004-506078.}

\end{document}